\documentclass[pre,twocolumn,a4paper,superscriptaddress,floatfix,10pt,nofootinbib,longbibliography]{revtex4-2}
\usepackage{graphicx}
\usepackage{dcolumn}
\usepackage{bm}
\usepackage{color}
\usepackage{amsmath}
\usepackage{amssymb}
\usepackage{hyperref}
\usepackage{algorithm}
\usepackage[end]{algpseudocode}

\newcommand{\h}{\mathit{\boldsymbol{h}}}
\newcommand{\X}{\mathit{\boldsymbol{x}}}

\newcommand{\loc}{\mathrm{loc}}

\newcommand{\beq}{\begin{equation}}
\newcommand{\eeq}{\end{equation}}
\newcommand{\ba}{\begin{array}}
\newcommand{\ea}{\end{array}}
\newcommand{\bea}{\begin{eqnarray}}
\newcommand{\eea}{\end{eqnarray}}

\begin{document}

\title{Simulating disordered quantum systems via dense and sparse restricted Boltzmann machines}

\author{S. Pilati}
\affiliation{School of Science and Technology, Physics Division, Universit{\`a}  di Camerino, 62032 Camerino (MC), Italy}

\author{P. Pieri}
\affiliation{School of Science and Technology, Physics Division, Universit{\`a}  di Camerino, 62032 Camerino (MC), Italy}
\affiliation{INFN, Sezione di Perugia, 06123 Perugia (PG), Italy}


\begin{abstract}
In recent years, generative artificial neural networks based on restricted Boltzmann machines (RBMs) have been successfully employed as accurate and flexible variational wave functions for clean quantum many-body systems.
In this article we explore their use in simulations of disordered quantum spin models.
The standard dense RBM with all-to-all inter-layer connectivity is not particularly appropriate for large disordered systems, since in such systems one cannot exploit translational invariance to reduce the amount of parameters to be optimized.
To circumvent this problem, we implement sparse RBMs, whereby  the visible spins are connected only to a subset of local hidden neurons, thus reducing the amount of parameters.  We assess the performance of sparse RBMs as a function of the range of the allowed connections, and compare it with the one of dense RBMs. 
Benchmark results are provided for two sign-problem free Hamiltonians, namely pure and random quantum Ising chains.
The RBM ansatzes are trained using the unsupervised learning scheme based on projective quantum Monte Carlo (PQMC) algorithms.
We find that the sparse connectivity facilitates the training process and allows sparse RBMs to outperform the dense counterparts.
Furthermore, the use of sparse RBMs as guiding functions for PQMC simulations allows us to perform PQMC simulations at a reduced computational cost, avoiding possible biases due to finite random-walker populations.
We obtain unbiased predictions for the ground-state energies and the magnetization profiles with fixed boundary conditions, at the ferromagnetic quantum critical point. The magnetization profiles agree with the Fisher-de Gennes scaling relation for conformally invariant systems, including the scaling dimension predicted by the renormalization-group analysis.
\end{abstract}

\maketitle

\section{Introduction}
\label{secintro}
Carleo and Troyer's 2017 article~\cite{carleotroyer} gave impetus to a vibrant research activity on the use of artificial neural networks as variational ansatzes for ground-state wave-functions.
Several neural networks have been adopted, but restricted Boltzmann machines (RBMs)~\cite{ackley1985learning,HintonConDiv,fischer2012introduction} have emerged as one of the most accurate and versatile~\cite{melko2019restricted}.
Beyond approximating ground-state wave functions, they have also been used to perform quantum state tomography~\cite{torlaitomography,torlai2019review}, to accelerate classical and quantum Monte Carlo simulations~\cite{huang2017accelerated,PhysRevE.100.043301}, to solve classical combinatorial optimization problems~\cite{gomes2019classical}, as well as to simulate excited states~\cite{choo2018symmetries}, open quantum systems~\cite{nagy2019variational,PhysRevLett.122.250502,PhysRevLett.122.250503,PhysRevB.99.214306}, and the unitary dynamics~\cite{carleotroyer}.
Recently, a procedure to represent via RBMs also states with  non-abelian symmetries has been introduced~\cite{PhysRevLett.124.097201}.
A plethora of competitive neural-network models have been investigated, including (possibly among others): unrestricted Boltzmann machines~\cite{inack3}, deep Boltzmann machines~\cite{carleo2018constructing}, deep feed-forward neural networks~\cite{lagaris1997artificial,saito,gao2017efficient,cai2018approximating,choo2018symmetries}, convolutional neural networks~\cite{saito2,liang2018solving,choo2019study}, generalized transfer matrix states~\cite{PhysRevB.99.165123}, neural network backflow models~\cite{luo2019backflow},  and neural Gutzwiller-projected wave functions~\cite{PhysRevB.100.125131}. 
More recently, autoregressive neural networks have also been employed~\cite{sharir2020deep,hibat2020recurrent,roth2020iterative}. 
Very different physical systems have already been simulated via artificial neural networks. The first studies addressed unfrustrated spin Hamiltonians~\cite{carleotroyer}. They have been followed by studies on bosonic lattice models~\cite{saito,saito2,cai2018approximating}, frustrated spin Hamiltonians~\cite{liang2018solving,choo2019study,szabo2020neural,duric2020efficient},  fermionic lattice models~\cite{luo2019backflow}, as well as on continuous-space bosonic~\cite{saito2018method,PhysRevLett.120.205302} and fermionic systems~\cite{PhysRevLett.120.205302}, in particular electronic ones~\cite{han2019solving,pfau2019ab,hermann2019deep,kessler2019artificial}. Topological states~\cite{deng2017machine,glasser2018neural,PhysRevB.99.155129} and nuclear systems have been addressed as well~\cite{keeble2019machine}.
The investigations mentioned above addressed few-body systems or clean extended models. To the best of our knowledge, Hamiltonians with random disorder have not been addressed so far. In particular, it is still unclear  whether neural-network wave functions can accurately describe the ground state of disordered quantum many-body systems.
%

\begin{figure}[h]
\begin{center}
\includegraphics[width=1.0\columnwidth]{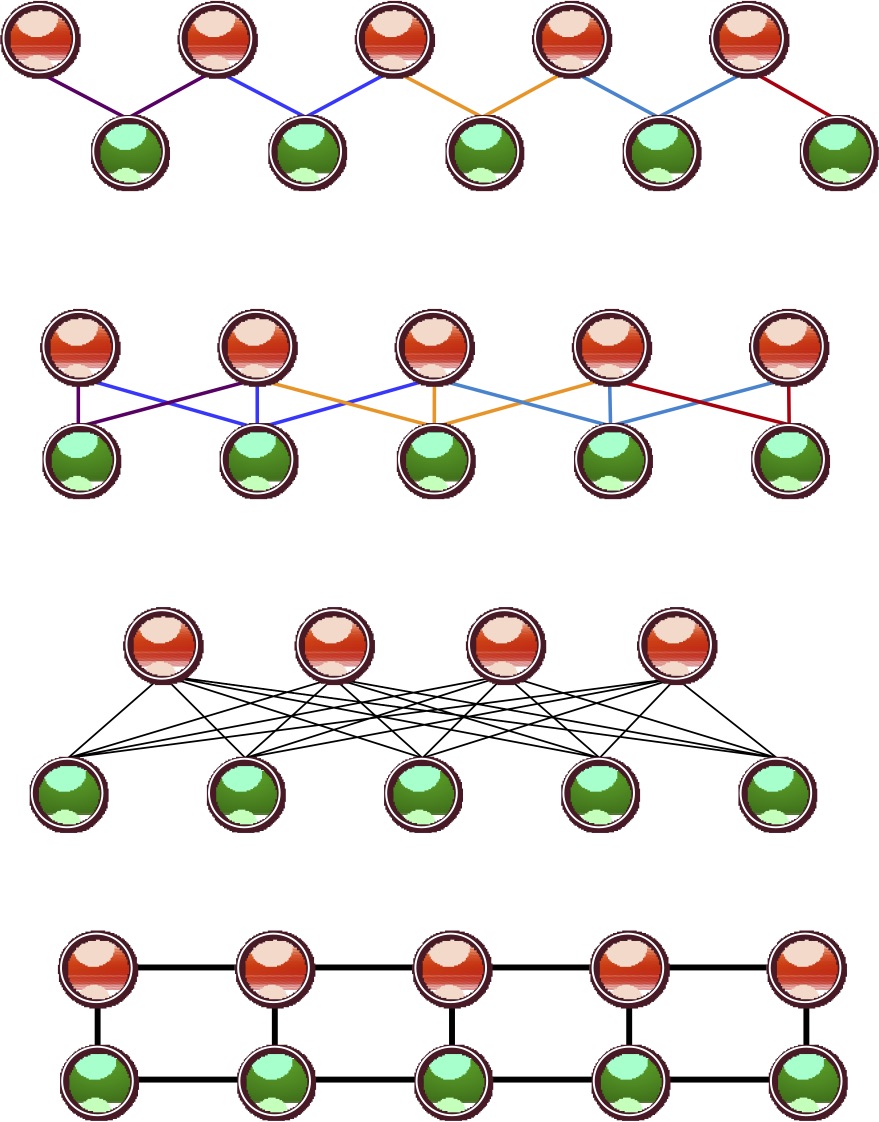}
\caption{(color online). Representation of the connectivity structures of four different artificial neural networks. Segments indicate the allowed interactions. 
The two upper diagrams represent examples of the sparse restricted Boltzmann machine employed in this article, whereby visible spins interact with a limited number of local hidden neurons; in the cases visualized here this number is $N_c=2$ (top) and $N_c=3$ (second from top). The third diagram from top represents the standard dense restricted Boltzmann machine with all-to-all inter-layer connectivity~\cite{carleotroyer}.
The bottom diagram represents a type of unrestricted Boltzmann machine~\cite{inack3}, alias shadow wave-function~\cite{reatto1988shadow,shadowvitiello}. This is characterized by inter-layer connections only among spins corresponding to the same index, plus nearest-neighbor intra-layer connections.
}
\label{fig1}
\end{center}
\end{figure}
%

In this article, we explore the use of artificial neural networks --- specifically, of RBMs --- to simulate the ground state of disordered quantum spin Hamiltonians. 
Among the three available schemes to train the RBMs, namely reinforcement learning~\cite{carleotroyer},  supervised learning~\cite{kochkov2018variational}, and unsupervised-learning~\cite{PhysRevE.100.043301}, we adopt the latter.
In this scheme, the training of the RBM is performed in combination with projective quantum Monte Carlo simulations (PQMC). On the one hand, this scheme provides one with accurate RBM approximations for the ground-state wave-function. On the other hand, the use of the optimized RBMs as guiding functions for successive PQMC simulations~\cite{PhysRevE.100.043301} allows one to  eliminate the possible bias due to the finite random-walker population~\cite{hetherington1984observations,boninsegnimoroni,inack2,inack3}. 
Notably, this provides one with unbiased estimates of ground-state properties, even when the optimized RBM wave-function is not exact.

Standard RBMs are characterized by a dense connectivity structure, whereby all visible spins are connected to all neurons in the (unique) hidden layer. In general, this implies that the number of parameters to be optimized increases with the square of the system size. 
Such a large amount of parameters often leads to difficulties in the optimization of variational ansatzes for many-body systems, meaning that the optimization algorithm might fail to identify optimal parameters. Furthermore, the dense connectivity increases the computational cost of any RBM-based Monte Carlo simulation, including PQMC simulations guided by RBMs.
In clean systems, translational invariance can be exploited to reduce the scaling of the number of parameters  with the system size from quadratic to  linear~\cite{sohn2012learning,carleotroyer}.
Clearly, this is not possible in disordered systems.
In this article, we investigate the use of sparse RBMs~\cite{deng2017machine,chen2018equivalence,glasser2018neural,sehayek2019learnability} for disordered systems. 
Specifically, a notion of spatial distance is embedded in the connectivity structure, and the visible spins are connected only to a certain number of nearest-neighbor hidden neurons.
As suggested by studies performed in the field of machine learning~\cite{mocanu2016topological,sehayek2019learnability}, one expects the sparse connectivity to allow the model learning the dominant correlations at a substantially reduced computational cost. This should also reduce  the risk of overfitting.
Notably, the sparse connectivity reduces the computational cost of evaluating wave-function ratios, substantially accelerating both variational and projective quantum Monte Carlo simulations.
The main testbed we consider in this article is the  random Ising chain, beyond the pure ferromagnetic model which has been addressed also in previous studies. 
We analyze the accuracy of sparse RBMs as a function of the range of the allowed connections, exploring both short-range and mid-range connectivities.  We also compare sparse RBMs  with their dense counterparts. Furthermore, the accuracy of PQMC simulations guided by optimized sparse RBMs is verified.
The observables we analyze are the ground-state energy  and the magnetization profile in setups with fixed boundary conditions.

Our analysis  indicates that the local connectivity facilitates the training process, allowing sparse RBMs to  approximate the ground-state of the random spin model more accurately than dense RBMs with a comparable number of variational parameters.
The PQMC simulations guided by optimized sparse RBMs provide unbiased results at a substantially reduced computational cost compared to the case of dense guiding functions.
In particular, the magnetization profiles computed by the PQMC algorithm satisfy the Fisher-de Gennes scaling relation~\cite{fisherdegennes}. The scaled profiles agree with the predictions for conformally invariant systems~\cite{burkhardt1985universal}, even for the (non-conformally invariant) random model, indicating that the corrections to the conformal scaling are essentially negligible, as previously found also in Ref.~\cite{igloi1997density}. The scaling dimension agrees with the results of the renormalization-group treatment~\cite{fisher1995critical}.

The rest of the article is organized as follows: Section~\ref{secmethod} introduces the model Hamiltonian as well as the dense and the sparse RBMs. The unsupervised-learning scheme and the PQMC algorithm are also briefly described.
Our results for ground-state energies and magnetization profiles are reported in Section~\ref{secresults}. They are compared with Jordan-Wigner predictions and with the Fisher-de Gennes scaling relation, respectively.
Our conclusions and some future perspectives are given in Section~\ref{secconclusion}.

\section{Models and Methods}
\label{secmethod}
The models we consider are defined by the following  one-dimensional quantum Ising Hamiltonian:
\begin{equation}
\hat{H}=-\sum_{j=1}^{N} J_j{\sigma}^{z}_{j} {\sigma}^{z}_{j+1} -\Gamma \sum_{j=1}^{N} {\sigma}^{x}_{j}.
\label{H}
\end{equation}
$\sigma^x_j$ and $\sigma^z_j$ are conventional Pauli matrices at the lattice sites $j=1,\dots,N$. $N$ is the total number of spins, and we consider periodic boundary conditions, i.e. $\sigma^{\alpha}_{N+1}=\sigma^{\alpha}_{1}$, for $\alpha=x,z$. The parameters $J_j\geqslant0$ fix the strength of the (ferromagnetic) interactions between the spins at the sites $j$ and $j+1$.
$\Gamma$ is the intensity of the (uniform) transverse magnetic field. 
In the following,  the eigenstates of the Pauli matrix ${\sigma^z_j}$ with eigenvalues $x_j=\pm1$ are denoted as $\left| x_j \right>$. 
%
%
The quantum states of $N$ spins $\left|\X \right> = \left| x_1 x_2 ... x_N\right>$, with $\X=(x_1,\dots,x_N)$, form the  computational basis considered in this article. With $\left| \psi \right>$ we denote the quantum state corresponding to the wave-function $\left< \X \right| \left. \psi \right> = \psi(\X)$.

If one chooses uniform couplings $J_j=J> 0$, for all $j=1,\dots,N$, the Hamiltonian~\eqref{H} describes the pure ferromagnetic Ising chain. 
This model undergoes a quantum phase transition from a paramagnetic phase for $\Gamma>J$ to a ferromagnetic phase for $\Gamma < J$. 
Beyond the pure model, we consider the random Ising chain with couplings $J_j$ sampled from a probability distribution $\mathcal{P}(J_j)$. 
One can always perform a gauge transformation in the Hamiltonian~\eqref{H} such that $J_j >0$ for $j=1,\dots,N$. So, we consider only distributions with support on positive couplings. 
Specifically, we consider the uniform distribution $\mathcal{P}(J)=\theta(J)\theta(1-J)$, where $\theta(x)$ is the unit step function: $\theta(x)=1$ for $x>0$ and $\theta(x)=0$ otherwise. We consider also the binary distribution $\mathcal{P}(J)=\delta(J-2)/2+\delta(J-1/2)/2$, where $\delta(x)$ is the Dirac delta function.
In the random Ising chain, the ferromagnetic quantum phase transition occurs at the critical transverse field $\Gamma=\exp(\overline{\ln(J)})$~\cite{shankar1987nearest}; the over-line indicates the average over the chosen probability distribution. In the case of the uniform distribution described above, one obtains $\Gamma\cong 0.36792$, while for the binary distribution the critical point is $\Gamma=1$.

\subsection{Dense and sparse restricted Boltzmann machines} 
As first shown in Ref.~\cite{carleotroyer}, the ground-state wave-function of a quantum spin Hamiltonian like \eqref{H} can be approximated  using  Boltzmann machines.  Ref.~\cite{carleotroyer} addressed pure quantum spin models via standard dense RBMs.
Here we address also random Ising models, using RBMs with dense and with sparse connectivities.
Boltzmann machines are generative stochastic neural networks commonly employed for density estimation, i.e. to infer the probability distribution underlying   a given (typically large) dataset of  samples.
RBMs are formed by the visible layer with the $N$ spin variables $\X$, and by a layer of hidden neurons, which includes $N_h$ additional spin variables $h_i=\pm 1$, with $i=1,\dots,N_h$. The set of hidden variables will be indicated as $\h=(h_1,\dots,h_{N_h})$.
The probability associated to each configuration $(\X,\h)$ of the two-layer system is written in the form of the Boltzmann weight $P(\X,\h)$ corresponding to a classical Ising Hamiltonian. 
RBMs are characterized by the absence of intra-layer visible-visible and hidden-hidden interactions. In the standard dense RBMs, all visible spins interact with all hidden spins. The corresponding coupling parameters are denoted $J_{ij}$. In the sparse RBMs, instead, each visible spin $j$ interacts only with a subset of hidden neurons. This subset is denoted in the following as $\mathcal{N}_j$.
The classical Ising Hamiltonian associated to the sparse RBM reads: 
\beq
\label{ham_rbm}
H_{{\rm RBM}} \left(\X,\h\right)=-\sum_{j}\sum_{i\in\mathcal{N}_j} J_{ij}h_i x_j -\sum_j a_j x_j - \sum_i b_i h_i.
\eeq
The parameters $a_j$ and $b_i$, called biases, play the role of local longitudinal magnetic fields. Together with the couplings $J_{ij}$, they define the RBM.
The parameters will be collectively indicates as ${\bf W}\equiv (\{J_{ij}\},\{a_{j}\},\{b_{i}\})$.
The dense RBMs correspond to the choice $\mathcal{N}_j=\{1,\dots,N_h\}$, for all $j=1,\dots,N$.
In the sparse RBM considered in this article,  the size of the hidden layer matches the one of the visible layer, i.e.~$N_h=N$. The visible spins are connected only to a number $N_c<N_h$ of nearest-neighbor hidden neurons, corresponding to the set of hidden-spin indices $\mathcal{N}_j=\{j-(N_c-1)//2,\dots,j-(N_c-1)//2+N_c-1\}$, where the symbol $//$  indicates integer division. Periodic boundary conditions are applied to the hidden-neuron indices, analogously to the visible-spin indices.
For dense RBMs, one has $N\times N_h$ inter-layer couplings $J_{ij}$, besides $N+N_h$ biases.
For sparse RBMs, the number of inter-layer couplings is $N\times N_c$ while the number of bias terms remains $N+N_h$.

The connectivity structures corresponding to the two architectures are visualized in Fig.~\ref{fig1}. Notice that the connections across the periodic boundary are not shown.
For comparison, it is also worth mentioning that in the case of the unrestricted Boltzmann machine considered in Ref.~\cite{inack3}, intra-layer visible-visible and hidden-hidden direct interactions are included, but they are limited to the nearest-neighbor couplings. The corresponding connectivity structure is also  shown in Fig.~\ref{fig1}.
This structure is analogous to the one of the shadow wave-function introduced in Refs.~\cite{reatto1988shadow,shadowvitiello} to describe the liquid and solid phases of Helium-4.

Sparse RBMs with short-range inter-layer connections have been previously considered also in Ref.~\cite{glasser2018neural}. In that reference, they have been show to be equivalent to a specific type of entangled plaquette states. Due to the (exponentially) more favorable scaling of the computation cost, they allow the use of larger plaquettes.~\cite{glasser2018neural}. It has also been shown that certain topological  states, namely the toric code and the one-dimensional symmetry-protected topological cluster state, admit an exact representation via RBM with local inter-layer connections~\cite{deng2017machine}.
In practical applications, accurately approximating the ground-state wave-function via dense RBMs requires a number of hidden units $N_h \gtrsim N$. Thus, the number of parameters to be optimized scales, to leading order (without counting the bias terms), at least as $N^2$. In clean periodic systems, translational invariance can be exploited to reduce this quadratic scaling to a linear scaling with system size~\cite{sohn2012learning,carleotroyer}. However, this is not possible in disordered models, which are the main focus of this article.
As shown in Section~\ref{secresults}, sparse RBMs can reach high accuracy already for $N_c \ll N$, implying a substantial reduction of the number of parameters to be optimized. 
It is worth mentioning that one could implement sparse RBM with larger hidden layers $N_h>N$. This would provide the same flexibility one has with dense RBMs. 
In this respect, it is worth recalling that dense RBMs with an arbitrarily large number of hidden neurons can, in principle, approximate any binary distribution~\cite{freund1992unsupervised,le2008representational}. They share this propriety with deep neural networks with many hidden layers.
 It follows that  a sparse RBM with sufficiently large $N_h$ and $N_c$ must also be a universal approximator. 
However, in practice the number of neurons in the unique hidden layer of an RBMs might have to exponentially increase with the system size~\cite{martens2013representational}, making  the training process and RBM-based Monte Carlo simulations computationally impractical.
This highlights the importance of identifying RBM architectures that reach a sufficient accuracy with a limited number of parameters.

The probability associated by the RBM to the visible-spin configuration $\X$ is computed as the marginal distribution over all possible hidden-spin configurations $\h$: 
\beq
P( \X ) = \sum_{\h }P(\X,\h) =  \frac{1}{Z}\sum_{\h } \exp\left[-H_{{\rm RBM}} ( \X,\h)\right].
\eeq
The normalization factor $Z = \sum_{\X,\h} \exp\left[ -H_{{\rm RBM}}  \left(\X,\h\right) \right]$ is the partition function.
The absence of intra-layer interactions allows one to analytically trace out the hidden-spin configurations,
resulting in the (unnormalized) marginal distribution 
$P(\X) \propto \exp\left(\sum_j a_j x_j\right) \prod_i F_i(\X)$, where $F_i(\X)=2\cosh\left[b_i + \sum_{j\in{\mathcal{N}_i}} J_{ij} x_j\right]$, and
$\mathcal{N}_i$ indicates the set of visible-spin indices $j$  connected to the hidden neuron $i$.
For the dense RBM, this set is $\mathcal{N}_i=\{ 1,\dots, N \}$ for any $i$.
For the sparse RBM we implement, one has $\mathcal{N}_i=\{ i-(N_c//2),\dots,i-(N_c//2)+N_c-1   \}$.
As proposed in Ref.~\cite{carleotroyer}, the function $P(\X)$ can be used to define an (unnormalized) ground-state wave-function, $\psi_{\mathrm{RBM}}(\X) \propto P(\X)$.

In the general case, the RBM parameters ${\bf W}$ should be complex valued in order to describe both amplitude and phase.
Otherwise, one could consider two real-valued RBMs, one for the amplitude and the other for the phase.
In this article we consider only models whose ground-state wave function can be assumed to be real and nonnegative in the chosen computational basis.
Therefore, the RBM parameters can be restricted to real values.
The RBM could be trained via the variational minimization of the energy expectation value for the state $\left| \psi_{{\rm RBM}}  \right>$~\cite{carleotroyer}.
This procedure is analogous to the reinforcement learning schemes employed in the field of machine learning, with the (negative) energy expectation-value playing the role of reward function.
In many machine learning applications of RBMs, the parameters ${\bf W}$  are instead determined via unsupervised machine-learning~\cite{ackley1985learning}. In these schemes, the RBM is trained to mimic the unknown distribution underlying a dataset of stochastic samples, as mentioned above.
To generate the training dataset, we use PQMC algorithms, as explained in the following.

\subsection{Projective Quantum Monte Carlo simulations}
PQMC algorithms are designed to simulate the ground-state of quantum systems by stochastically evolving the imaginary-time Schr\"odinger equation. 
In the practical implementation, the algorithm efficiency is boosted by introducing  a suitable ansatz for the ground-state wave-function. This is usually referred to as guiding function. It is indicated in the following as $\psi_G({\X})$. In the PQMC simulation, one lets evolve the product  $f({\X},\tau)=\psi ({\X},\tau) \psi_G({\X})$, where $\psi ({\X},\tau)$ is the time-dependent wave-function, according to the modified imaginary-time Schr\"odinger equation:
\begin{equation}
\label{masterf}
f(\X,\tau+\Delta \tau) = \sum_{\X^\prime} \tilde{G}(\X,\X^\prime,\Delta \tau) f(\X^\prime,\tau). 
\end{equation}
Here, $\tilde{G}({\X},{\X}^\prime,\Delta \tau)= G(\X,\X^\prime,\Delta \tau)\frac{\psi_G({\X})}{\psi_G({\X^\prime})}$, 
where $G({\X},{\X}^\prime,\Delta \tau)= \left< \X \left| \exp\left[{-\Delta \tau(\hat{H}-E_{\mathrm{ref}})}\right] \right|\X^\prime \right>$ 
is the imaginary-time Green's function for a time step $\Delta \tau$,  $E_{\mathrm{ref}}$ is a reference energy introduced to stabilize the dynamics, and we set $\hbar =1$ throughout this article.

The dynamics is simulated by stochastically evolving a large population of system configurations, usually called random walkers. 
These random walkers are subjected to stochastic configuration updates and to a branching process whereby they are replicated or annihilated. 
%
%
Specifically, we employ the continuous-time algorithm of Refs.~\cite{becca_sorella,SorellaCTGFMC,inack3}. This avoids finite time-step errors.
In order to sample the configuration updates one has to compute the ratio of guiding-function values $\frac{\psi_G({\X})}{\psi_G({\X^\prime})}$, for all possible transitions $\X^\prime\rightarrow \X$ induced by the Hamiltonian. 
This constitutes the computational bottleneck of PQMC simulations.
In the case of~\eqref{H}, the allowed transitions correspond to the $N$ possible spin flips. 
If an RBM wave-function is chosen as guiding function, i.e. if $\psi_G({\X})=\psi_{\mathrm{RBM}}(\X)$, the ratio for the flip of spin $j$ is evaluated as:
\begin{equation}
\label{ratio}
\frac{\psi_G({\X})}{\psi_G({\X^\prime})}=\exp(-2a_jx_j^\prime)\prod_{i\in\mathcal{N}_j} \frac{F_i(\X)}{F_i(\X^\prime)},
\end{equation}
where $\X=\{x_1^\prime,\dots,-x_j^\prime,\dots,x_N^\prime\}$.
Due to the reduced number of factors,  sparse RBMs with $N_c \ll N_h$ lead to a substantial reduction of the computational cost of evaluating Eq.~\ref{ratio}. One obtains a speed-up of order $N_h/N_c$ in the PQMC simulation.
It is worth mentioning that, for both dense and sparse RBMs, the computational cost of evaluating the arguments $b_i+\sum_{j\in\mathcal{N}_i}J_{ij}x_j$ of the $\cosh$ function appearing in $F_i(\X)$ can be substantially reduced by book-keeping their values, updating them only when a visible-spin flip is performed.
 To implement the branching process and for the tuning of $E_{\mathrm{ref}}$ we use the textbook recipe of Ref.~\cite{thijssen}.
The tuning of $E_{\mathrm{ref}}$ is designed to stabilize the average number of random walkers close to a target value $N_{\mathrm{w}}$.
In the long imaginary-time limit, the random walkers sample spin configurations with a probability distribution proportional to $f(\X,\tau\rightarrow \infty) \cong\psi_G({\X})\psi_0({\X})$, where $\psi_0({\X})$ is the ground-state wave function.
In the large random-walker population limit, $N_{\mathrm{w}}\rightarrow \infty$, the above approximate identity becomes exact. Therefore, one obtains unbiased estimates of the ground-state energy $E$ via Monte Carlo integration of the following formula:
\beq
\label{energyPQMC}
E = \frac{\sum_{\X} f\left(\X,\tau\rightarrow\infty\right)E_{\mathrm{loc}}(\X)}{ \sum_{\X} f\left(\X,\tau\rightarrow\infty\right)},
\eeq
where the local energy is $E_{\loc}(\X')=\sum_{\X} H_{\X',\X}\frac{\psi_G({\X})}{\psi_G({\X^\prime})} $, with $H_{\X',\X}=\left< \X \left| \hat{H} \right|\X^\prime \right>$.
Analogous formulas can we written for other operators that commute with the Hamiltonian. The forward-walking technique provides unbiased estimates also for other operators with a diagonal representation in the chosen computational basis (see, e.g., Ref.~\cite{boronat}).
For finite $N_\mathrm{w}$ a systematic bias might occur~\cite{hetherington1984observations,golinelli1992haldane,nemec,boninsegnimoroni,inack2,pollet2018stochastic}. 
This is known to be the major deficiency of PQMC algorithms.
In fact, if one does not introduce a guiding function, i.e., if one sets  $\psi_G({\X})=1$, the random-walker population $N_{\mathrm{w}}$ required to keep this systematic bias below a given threshold increases exponentially with the system size~\cite{inack2}. 
If $\psi_G({\X})$ is a reasonably good approximation for the ground-state wave function $\psi_0({\X})$ the convergence to the exact $N_{\rm w}\rightarrow \infty$ limit is drastically accelerated~\cite{inack3}. 
This allows one to reduce the bias to negligible values, thus obtaining unbiased predictions with a computationally affordable random-walker population.
As shown in Ref.~\cite{PhysRevE.100.043301} and briefly summarized below, one can use configurations sampled from PQMC simulations to train RBM wave-functions in an unsupervised learning scheme. The trained RBMs can be used in turn as guiding function for a subsequent PQMC run,  eliminating the finite $N_{\mathrm{w}}$ bias. This scheme is briefly described in subsection~\ref{selfPQMC}.

\subsection{Unsupervised learning for RBMs}
In unsupervised learning, the RBM is trained by maximizing the log-likelihood $L(\bold{W}) = \sum_l \ln P(\X_l)$ of a training set;  the index $l$ labels the instances in the training set. 
This is equivalent to the minimization of the 
Kullback-Leibler divergence~\cite{fischer2012introduction}. For two generic distributions $p(\X)$ and $q(\X)$, the Kullback-Leibler divergence reads:
\begin{equation}
\mathrm{KL} \left( q \left| \right| p\right) = \sum_{\X} q(\X) \ln\frac{q(\X)}{p(\X)}.
\end{equation}
In the case discussed here, $q(\X)$ is identified with the distribution of the random walkers $f(\X,\tau\rightarrow \infty)$ obtained via PQMC simulations, while $p(\X)$ corresponds to the RBM marginal distribution $P(\X)$.
The optimization of the RBM parameters $\bf{W}$  can be performed using the stochastic gradient ascent algorithm. One performs many updates of the RBM parameters following the log-likelihood gradient computed  on small, randomly sampled, mini-batches of training instances.
The gradients are computed via the $k$-step contrastive divergence algorithm~\cite{bengio2009justifying}. This involves $k$ iterations of alternated Gibbs sampling of hidden and visible variables, starting from the visible values of the training instances.
In this article, the plain vanilla stochastic gradient ascent algorithm is augmented only by adding a momentum term proportional to the gradient computed at the previous update.
Also learning-rate annealing is adopted, as explained below.
All details of this algorithm are described in Ref.~\cite{fischer2012introduction}. Notice that, for the values of the  binary spin-variables, we follow the physics convention $x_j=\pm1$ and $h_i=\pm 1$, instead of the convention $x_j=\{0,1\}$ and $h_i=\{0,1\}$, which is more common in the machine-learning literature.
The modified algorithm corresponding to the physics notation is provided in Ref.~\cite{PhysRevE.100.043301}. See also Refs.~\cite{huang2017accelerated,torlai2019review}.
The use of sparse RBMs implies simple algorithmic modifications with respect to the standard dense models. In particular, one has to compute only $N\times N_c$ gradients with respect to the coupling parameters, as opposed to  $N\times N_h$. 
Gibbs sampling is performed with the following binary probability distributions: 
\begin{equation*}
p_{h_i=1}(\X) = \frac{1}{ 1 + \exp\left( -2\sum_{j\in\mathcal{N}_i} x_j J_{ij} -2b_i \right) },
\end{equation*}
and
\begin{equation*}
p_{x_j=1}(\h) = \frac{1}{ 1 + \exp\left( -2\sum_{i\in\mathcal{N}_j} h_i J_{ij} -2a_j \right) },
\end{equation*}
for hidden and visible variables, respectively. Clearly, $p_{h_i=-1}(\X)=1-p_{h_i=1}(\X)$, and $p_{x_j=-1}(\h)=1-p_{x_j=1}(\h)$.
%
%
%

\begin{figure}
\begin{center}
\includegraphics[width=1.0\columnwidth]{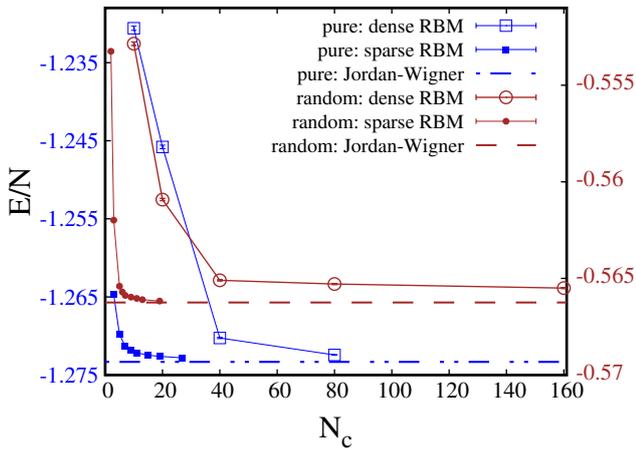}
\caption{(color online). 
Energy per spin $E/N$ corresponding to two optimized neural-network ansatzes, as a functions of the number of connected hidden spins $N_c$. For the dense RBMs, which are represented by the empty symbols, $N_c$ coincides with the number of hidden spins $N_h$. 
The full symbols correspond to the sparse RBMs, for which $N_c$ is the number of allowed local connections. 
The (blue) squares represent the results for the clean ferromagnetic Ising model, and they refer to the left vertical axis. The (brown) circles correspond to one realization of the random Ising model, and they refer to the right vertical axis. Both models include $N=80$ spins and they are simulated at the coupling strength corresponding to the respective ferromagnetic quantum critical points.
The horizontal lines indicate the exact ground-state energies computed through the Jordan-Wigner transformation for $N=80$.
}
\label{fig2}
\end{center}
\end{figure}

\subsection{Self-learning PQMC algorithm}
\label{selfPQMC}
Ref.~\cite{PhysRevE.100.043301} presented a self-learning PQMC algorithm whereby an unsupervised learning scheme is used to train the RBM.
Specifically, the RBM learns the probability distribution corresponding 
to a large random-walker population produced by a PQMC simulation after a long imaginary-time projection. 
The self-learning protocol is actually implemented by performing several consecutive PQMC runs, hereby labelled by the index $s$, for an imaginary-time $\tau_s$. Each run is guided by a different guiding function $\psi_{G_s}(\X)$, and it is followed by an unsupervised learning stage which provides an RBM such that $P_s(\X) \propto \psi_{G_s}(\X)\psi_0(\X)$. The guiding function for the next PQMC run is $\psi_{G_{s+1}}(\X)=\sqrt{P_{s}(\X)}$. 
For the initial run, a guiding function based on the square root of an RBM with random parameters is used.
The use of the square root of the RBM probability distribution implies straightforward modifications to the formulas for the guiding-function ratios~Eq.~\eqref{ratio} and for $E_{\mathrm{loc}}(\X)$.
If $N_{\rm w}$ and  $\tau_s$ are large enough, and if the training of the RBM succeeds, one has a fast convergence $\psi_{G_{s}}(\X)\rightarrow \psi_0(\X)$ for $s\rightarrow\infty$~\cite{PhysRevE.100.043301}.
Even if these assumptions are not exactly satisfied, this self-learning scheme provides remarkably accurate approximations for the ground-state wave function.
This accuracy is quantified in Section~\ref{secresults} by comparing the energy expectation value corresponding to the optimized RBM guiding function, defined as
\beq
\label{energyRBM}
E  =  \frac{\langle \psi_{G_{s\rightarrow \infty}} | \hat{H} |  \psi_{G_{s\rightarrow \infty}}\rangle} {\langle \psi_{G_{s\rightarrow \infty}} |  \psi_{G_{s\rightarrow \infty}} \rangle},
\eeq
with the exact ground state energy.
 These expectation values can be determined via standard Monte Carlo integration performed with the single spin-flip Metropolis algorithm. 
Even  more importantly, the RBM guiding function turns out to be  sufficiently accurate to eliminate the possible bias in the PQMC results due to the finite $N_{\rm w}$.
This is demonstrated by the data reported in Section~\ref{secresults}.

\subsection{Simulation details}
The results reported in Section~\ref{secresults} are obtained with the following simulation parameters.
The target random-walker population is $N_{\rm w}=10^4$.  The time step is $\Delta \tau=0.04$.
The final imaginary time of each PQMC runs is $\tau_s=20$.
The training dataset is accumulated by storing $N_{\rm w}/20$ randomly-selected walkers at each imaginary-time step, excluding the initial time segment of each PQMC run corresponding to $\tau\in\left[20s,20s+8\right]$.  The number of PQMC runs ranges from $20$ to $50$.
In each unsupervised-learning stage, the number of stochastic gradient ascent steps is $N_{\mathrm{steps}}=5\times 10^4$. The mini-batch size is $N_b=20$. The learning rate $\eta$ is kept fixed within each learning stage, but it is annealed stage after stage,  following the simple empirical protocol $\eta(s)= \eta_0 c^s$; the annealing rate is between $c=0.65$ and $c=0.85$, and  the learning rate at the first learning training stage $s=0$ is $\eta_0 = 0.01$.
The coefficient of the momentum term is $\nu=\eta/10$.
The one-step contrastive divergence algorithm, corresponding to $k=1$, is found to suffice.
The initial guiding function $\psi_{G_{s=0}}(\X)$ is the square root of an RBM with uniform random couplings in the range $J_{ij}\in \left[-0.025:0.025\right]$. 
For the simulations performed in  the absence of  a longitudinal magnetic field at the boundaries,  the bias terms $a_j$ and $b_i$ are initialized to zero, and they are not updated during the training processes. This choice reflects the spin-flip symmetry of the ground state.
For the simulations performed with a nonzero longitudinal field at the boundaries, the bias terms are initialized  to uniform random values $a_j,b_i \in \left[0:0.05\right]$. They are then optimized during the training stage.

\section{Results}
\label{secresults}
To analyze the accuracy of the optimized RBM wave functions, we compare the corresponding energy expectation value \eqref{energyRBM} with the exact ground-state energy. The comparison is performed for the pure and the random ferromagnetic quantum Ising chains. In the latter case, the couplings are sampled from the uniform distribution defined in Section~\ref{secmethod}. 
The exact ground-state energies, hereafter denoted with $E_{\mathrm{JW}}$, are determined via the Jordan-Wigner transformation. We follow the formalisms reported in Refs.~\cite{pfeuty1970one,cabrera1987role} and in Ref.~\cite{young1996numerical}, for the pure and the random models, respectively.
The comparison  for the system size $N=80$ is shown in Fig.~\ref{fig2}. 
This analysis is performed at the respective quantum critical points for the two models (see Section~\ref{secmethod}). The motivation is that, in this regime, it is more challenging to approximate ground-state wave-functions with neural-network ansatzes~\cite{carleotroyer,inack3,glasser2018neural,collura2019descriptive}.
%
%
\begin{figure}
\begin{center}
\includegraphics[width=1.0\columnwidth]{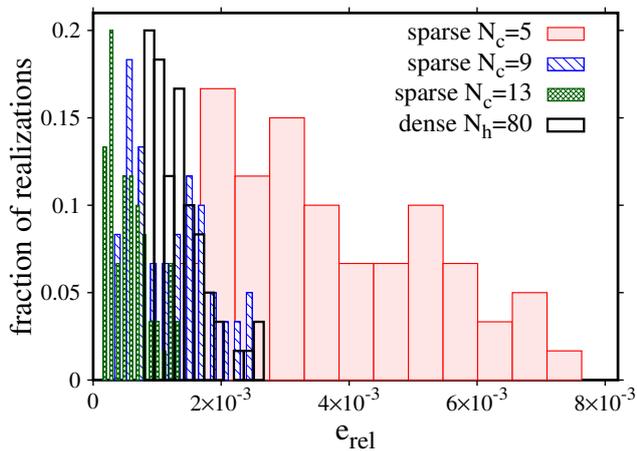}
\caption{(color online). 
Histogram of the relative errors $e_{\mathrm{rel}}=(E-E_{\mathrm{JW}} )/\left|E_{\mathrm{JW}} \right|$ of the ground-state energies $E$ corresponding to the optimized RBM ansatzes, with respect to the (exact) Jordan-Wigner result $E_{\mathrm{JW}} $. $60$ realizations of the uniform random couplings are considered.
The transverse field intensity is set at the ferromagnetic quantum critical point. 
The full columns correspond to three sparse RBMs different connection numbers $N_c$.
The empty columns correspond to a dense RBM with $N_h=80$ hidden neurons.
 The width of some columns is reduced for visibility.
}
\label{fig3}
\end{center}
\end{figure}
For the pure model, the dense RBM reaches a relative error $e_{\mathrm{rel}}=(E-E_{\mathrm{JW}})/|E_{\mathrm{JW}}| \simeq 7\times 10^{-4}$ with $N_h=80$ hidden neurons. 
The sparse RBM reaches a similar  relative error with just $N_c=15$ allowed connections per visible spins, while with $N_c=27$ one has  $e_{\mathrm{rel}}\simeq 4\times 10^{-4}$
For the specific realization of the random chain considered in Fig.~\ref{fig2}, the dense RBM is not particularly accurate, reaching $e_{\mathrm{rel}} \simeq 1.3\times 10^{-3}$ with $N_h=160$. Training even larger dense RBMs becomes computationally challenging.
Instead, the sparse RBM with $N_c=19$ provides $e_{\mathrm{rel}}\simeq 1.3\times 10^{-4}$, corresponding to an accuracy improvement of one order of magnitude.
These findings suggest that sparse RBMs should be preferred to the standard dense RBMs, since for pure models they achieve comparable accuracy at a reduced computational cost and, most importantly, they outperform dense RBMs for random models. This is due to the better optimization process that can be attained with a reduced number of variational parameters.
%

\begin{figure}[t]
\begin{center}
\includegraphics[width=1.0\columnwidth]{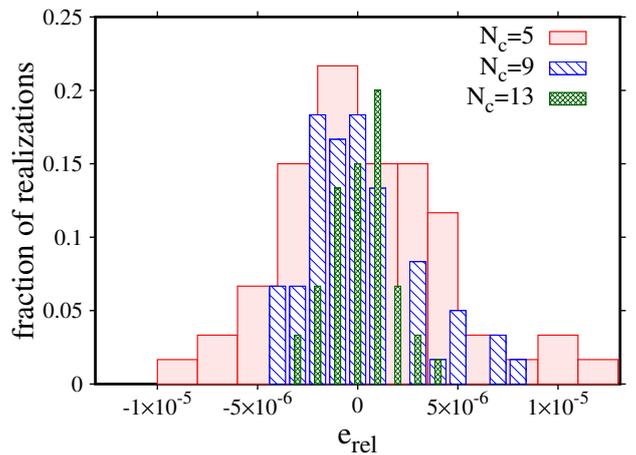}
\caption{(color online). 
Histogram of the relative errors $e_{\mathrm{rel}}$ of the ground-state energies $E$ predicted by projective quantum Monte Carlo simulations guided by optimized sparse RBMs. $60$ realizations of the uniform random couplings of the quantum Ising chain are considered.
The system parameters are as in Fig.~\ref{fig3}.
The three datasets correspond to different connection numbers $N_c$.
 The width of some columns is reduced for visibility.
}
\label{fig4}
\end{center}
\end{figure}
With disordered systems, it is crucial to verify that high accuracy can be achieved with any realization of the random couplings. 
To verify this, we analyze in Fig.~\ref{fig3} the histogram of the relative errors for $60$ realizations. The four datasets correspond to three sparse RBMs with different $N_c$, and to one dense RBM with $N_h=80$. For the sparse RBMs, one observes a systematic improvement of the accuracy with increasing $N_c$, for both average and maximum errors. 
For the dense RBM, the average  relative error is $\overline{e}_{\mathrm{rel}} = 1.4(4)\times 10^{-3}$ (the standard deviation is in parenthesis), which is comparable to the result for the sparse RBM with $N_c=9$, namely $\overline{e}_{\mathrm{rel}}  =1.2(6)\times 10^{-3}$.
This is remarkable, given that the sparse RBM has only a fraction $N_c/N_h$ of the variational parameters included in the dense RBM.
This confirms  that the sparse RBMs are particularly suitable to describe random spin systems.

One of our main goals is to obtain sufficiently accurate guiding functions to eliminate the bias in the PQMC simulations due to the finite random-walker population.
Fig.~\ref{fig4} shows the histogram of the energy predictions from PQMC simulations guided by the optimized sparse RBMs. These predictions are computed via Monte Carlo integration of Eq.~\eqref{energyPQMC}. Again, $60$ realizations of the random Ising chain are considered and the analysis is performed at criticality, since this is the regime where eliminating the bias requires larger random-walker populations~\cite{inack2,inack3}.
The PQMC results precisely agree with the Jordan-Wigner predictions, displaying only random fluctuations around $e_{\mathrm{rel}}=0$. These fluctuations are compatible with the statistical uncertainty of the Monte Carlo integration (not shown in the figure). As expected, these fluctuations decrease for increasing $N_c$, since the guiding function becomes more accurate. This reduces the fluctuations in the local energy $E_{\mathrm{loc}}(\X)$ and, therefore, in the result of the Monte Carlo integration.
%

\begin{figure}[t]
\begin{center}
\includegraphics[width=1.0\columnwidth]{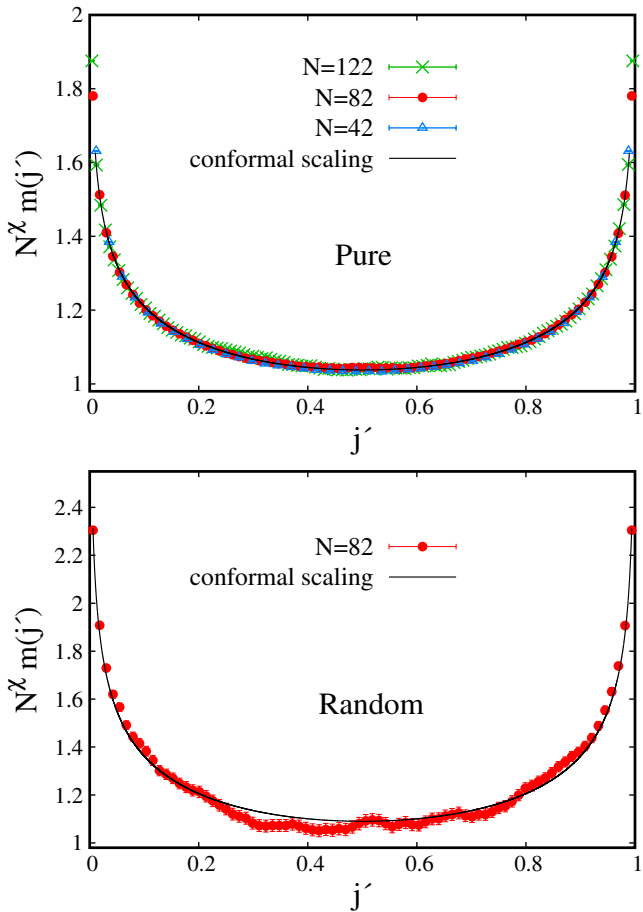}
\caption{(color online). 
Profile of the rescaled local magnetization $N^{\chi}m(j)$ as a function of the rescaled spin index $j\prime=(j-1/2)/N$, where $j=1,\dots,N$ is the site index and $N$ is the system size. $m(j)$ is the average magnetization at the site $j$ and $\chi$ is the scaling dimension of the magnetization. 
The different symbols represent the PQMC predictions for different $N$.
The system has fixed boundary conditions $m(1)=m(N)=1$. 
The transverse field intensity corresponds to the ferromagnetic quantum critical point.
The upper panel corresponds to a pure ferromagnetic Ising model.
The lower panel corresponds to a random Ising chain, with the couplings sampled from a binary distribution (see text).  The results are averaged over $1200$ realizations.
 The continuous (black) curves represent fitting functions based on the Fisher-de Gennes scaling relation for conformally invariant models; see Eq.~\eqref{scaling}.
The fitted scaling dimensions are consistent with the  expected results, $x_m=1/8$ and $x_m=(3-\sqrt{5})/4$, for the pure and the random models, respectively (see text).
}
\label{fig5}
\end{center}
\end{figure}
%

As discussed above, the sparse RBMs allow us to perform efficient PQMC simulations of random and, more in general, of inhomogeneous quantum spin systems. As a further showcase, we consider the inhomogeneity due to fixed boundary conditions. In this setup, the average magnetizations of the extremal spins are fixed at their maximum values, i.e. $m(j)\equiv\left<\psi_0\right| \sigma^z_j\left|\psi_0\right>/\left<\psi_0\right|\left.\psi_0\right>=1$, for $j=1$ and $j=N$.
The fixed boundary conditions are conveniently implemented by applying strong longitudinal magnetic fields on the extremal spins. Specifically, we add to the Hamiltonian~\eqref{H} a border term $-\lambda(\sigma^z_1+\sigma^z_N)$ with $\lambda=500$.
Notice that the fixed boundary conditions break the spin-flip symmetry of the Hamiltonian.
We determine via PQMC simulations the average magnetization profiles $m(j)$ of  the pure ferromagnetic model and of a random chain with couplings sampled from the binary distribution defined in Section~\ref{secmethod}. The guiding function is the optimized sparse RBM with $N_c=9$.
The PQMC predictions are compared with the Fisher-de Gennes scaling theory $m(j)=N^{-\chi} f(j/N)$, where $\chi$ is the scaling dimension of the magnetization operator and $f(j/N)$ is its scaling function.
For conformally invariant two-dimensional models (that is, 1+1, in the present quantum model), and assuming equal boundary conditions at the two extremes, as we do here, the scaling relation can be written as:
\begin{equation}
\label{scaling}
m(j)=A\left[\frac{N}{\pi}\sin\left(\pi \frac{j}{N}\right)\right]^{-\chi},
\end{equation}
where $A$ is a nonuniversal prefactor~\cite{burkhardt1985universal}.
More in general, Eq.~\eqref{scaling} represents the first term of a Fourier expansion~\cite{igloi1997density}.
In the upper panel of Fig.~\ref{fig5}, the PQMC prediction for the pure model is compared with the conformal-invariance scaling. The agreement is very precise. This is expected, since the pure model is conformally invariant. By using the scaling dimension $\chi$ as a fitting parameter on the $m(j)$ data for $N=122$, besides the prefactor $A$, we obtain $\chi=0.131(5)$, in agreement with the expected result $\chi=1/8$.
The error bar takes into account the fluctuations observed when removing up to three extremal points from the fit.
The random Ising chain breaks conformal invariance. However, it was found in Ref.~\cite{igloi1997density} that the Fourier terms beyond the first one provide  almost negligible contributions. In the lower panel of Fig.~\ref{fig5}, the conformal scaling is compared with the PQMC results averaged over $1200$ realizations of the random couplings. The agreement confirms the finding of  Ref.~\cite{igloi1997density}. 
By fitting the conformal scaling relations to the PQMC results we find $\chi=0.19(1)$, in agreement with the renormalization-group prediction for random Ising chains $\chi=(3-\sqrt{5})/4\cong0.19098$~\cite{fisher1992random,fisher1995critical}.

\section{Conclusions}
\label{secconclusion}
We have investigated the use of artificial neural networks to approximate the ground-state wave-function of disordered quantum spin models.
Our focus was on the feasibility of reaching high accuracy and on the computational efficiency of the neural-network models. Both issues are substantially more relevant in quantum many-body physics compared to the typical applications of neural networks in computer science and in engineering.
We have found that restricted Boltzmann machines with a local sparse connectivity reach higher accuracy, when trained via unsupervised learning, compared to the standard dense RBMs with all-to-all inter-layer connectivity. 
This is a consequential finding that highlights the crucial role of the connectivity structure of the neural-network wave-functions. 
The reduced connectivity of the sparse RBMs we implemented leads to a linear scaling with system size of the number of optimizable parameters. This has to be compared to the generally quadratic scaling of dense RBMs. The sparse connectivity facilitates the training process, allowing the RBM-model to better approximate the ground state, and it also accelerates the Monte Carlo simulations of RBMs, since the computational cost to evaluate wave-function ratios is reduced compared to the case of dense connectivity.
The optimized RBMs can also be used as guiding functions for projective quantum Monte Carlo simulations. In particular, we have shown that they allow one to completely eliminate the bias due to the finite random walker population in disordered spin systems. 
This possible bias is a possible weakness of the PQMC algorithms~\cite{ceperley1986quantum,foulkes2001quantum}, and the lack of appropriate guiding functions has so-far limited the scope of their application in this field.
Notably, the sparse connectivity  accelerates also the PQMC simulations.

As a future perspective, sparse RBMs could find use in studies of combinatorial optimization problems. In that context, projective quantum Monte Carlo algorithms have emerged as a stringent benchmark for physical quantum annealers~\cite{Polkovnikov_PRL15,PhysRevB.87.174302,inack2,PhysRevB.100.214303}. However, the lack of  guiding functions appropriate for the typical instances of complex optimization problems, which can be mapped to spin-glass models, has limited their success~\cite{santoroGFMC}.
It would also be interesting to investigate different sparse connectivities inspired by the theory of complex networks. Interesting candidates are RBM with  scale-free or small-world topologies~\cite{mocanu2016topological}.
We leave these endeavors to future studies.

\section*{Acknowledgements}
\noindent
S. P. and P. P. acknowledge financial support from the FAR2018 project titled ``Supervised machine learning for quantum
matter and computational docking'' of the University of Camerino and from the Italian MIUR under the project PRIN2017 CEnTraL 20172H2SC4.
S. P. also acknowledges the CINECA award under the ISCRA initiative, for the availability of high performance computing resources and support.

\bibliography{Ref}

\end{document}